# Molecular Dynamics Study of Stiffness in Polystyrene and Polyethylene

Mahdi Ahmadi Borji
*Emails: mehdi.p83@gmail.com, borji@webmail.guilan.ac.ir*

*Department of Physics, Institute for Advanced Studies in Basic Sciences, Gava Zang, Zanjan 45195-1159, Iran*
*Department of Physics, University of Guilan, Rasht, Iran*

**Abstract:**
In this paper, we have studied polystyrene (PS) and polyethylene (PE) stiffness by 3-dimensional Langevin Molecular Dynamics simulation. Hard polymers have a very small bending, and thus, their end-to-end distance is more than soft polymers. Quantum dot lasers can be established as colloidal particles dipped in a liquid and grafted by polymer brushes to maintain the solution. Here by study on molecular structures of PS and PE, we show that the principle reason lies on large phenyl groups around the backbone carbons of PS, rather than a PE with Hydrogen atoms. Our results show that the mean radius of PS random coil is more than PE which directly affects the quantum dot maintenance. In addition, effect of temperature increase on the mean radius is investigated. Our results show that by increasing temperature, both polymers tend to lengthen, and at all temperatures a more radius is predicted for PS rather than PE, but interestingly, with a difference in short and longer chains. We show that stiffness enhancement is not the same at short and long polymers and the behavior is very different. Our results show a good consonance with both experimental and theoretical studies.

**Introduction**

Polymers are molecules with a very large size with many atoms joining together in the form of a chain, and that is why they have high molecular weight and properties different from common small molecules in organic and inorganic chemistry. For example, PS is one of the polymers with many applications in industry. It is brittle and rigid. To remove this feature, it is made by copolymerization and in blends. Many behaviors of PS arise from its chemical structure. Its mechanical properties arise from presence of phenyl groups around backbone carbons [1-4]. It is shown that PS is harder than PE, and their $t_g$ are approximately $100^oC$, and $-120^o C$ respectively. Since PS contains many large phenyl groups rather than small hydrogen atoms in PE, Carbons of the backbone cannot choose all the directions to go. Therefore, movement and flexibility of PS is less than PE [5-8]. Nowadays, in quantum dot technology, polymers play a very instructive role. Maintenance of quantum dot colloids of CdS can be enhanced by making a polymer brush shell over them. The polymer shell can be efficient if it is made of a hard polymer with lower shrinkage. Thus, we are interested in polymers which can stay with less piling [15].

At zero temperature, with no external effects, polymers seem rod-like with no bending. Polymers usually are not in their ground state energy with configurations which are familiar. Firstly, in room temperature, the configuration breaks down, and remains only a stochastic behavior.

Another problem is that, polymers are mostly entangled into each other in a solid, or under stochastic forces due to solvent molecules' random motions.

**Our model:**

To simulate a polymer, we have used a 3-dimensional molecular dynamics simulation. The Langevin equation of motion for a particle is

$$m \ddot{r}_d = -\frac{\partial \mathcal{V}(r_d)}{\partial r_d} - \mu \dot{r}_d + \xi_d(t) \quad (1)$$

In which $m$ is the mass, $r_d$ represents $x$, $y$ and $z$, and the dot shows a derivation in term of time, $\mathcal{V}(r_d)$ potential over the particle, $\mu$ is friction coefficient, and $\xi_d$ are mutually uncorrelated Gaussian white noises appear due to thermal noises in the system, and obey the fluctuation-dissipation relation

$$<\xi_d(t)\xi_{d'}(t')> = 2\mu k_B \mathcal{T}\delta_{dd'}\delta(t-t') \quad (2)$$

In which $k_B$ is the Boltzmann constant and $\mathcal{T}$ is temperature of the system [9]. By the following changes in the variables

$$T = k_B\frac{\mathcal{T}}{\mathcal{V}_0}, \quad r_d = \frac{r_d}{L}, \quad V(r_d) = \frac{\mathcal{V}(r_d)}{\mathcal{V}_0},$$

$$\gamma = \frac{\mu L}{\sqrt{m\mathcal{V}_0}}, \quad t = t\sqrt{\frac{\mathcal{V}_0}{mL^2}} \quad (3)$$

in which $\mathcal{V}_0$ is the maximum amount of potential and $L$ is the characteristic length of the system. The dimensionless equations can be obtained as

$$\ddot{r}_d = -\frac{\partial V(r)}{\partial r_d} - \gamma \dot{r}_d + \xi_d(t) \quad (4)$$

where

$$<\xi_d(t)\xi_{d'}(t')> = 2\gamma T\delta_{dd'}\delta(t-t') \quad (5)$$

**IV. DETAILS OF OUR NUMERICAL METHOD**

In the Langevin Molecular Dynamics simulation, firstly, a system of N particles is constructed with an initial position for each of them. Then, we let particles interact with a Lenard-Jones potential [1], [2]. The potential for $i$th particle due to presence of $j$th one at distance $r_{ij}$ is

$$V_{LJ} = \varepsilon\left[\left(\frac{r_0}{r}\right)^{12} - 2\left(\frac{r_0}{r}\right)^6\right] \quad (6)$$

Which is attractive for particle interactions. However, if the distance is smaller than bond-length of $r_0$, it is repulsive. The Lenard-Jones force is calculated as

$$f_d = -\frac{\partial V_{LJ}}{\partial d} \quad (7)$$

In addition, to form a polymer, a spring potential must be taken into account between the links [3]:

$$U_{ij} = \frac{1}{2}K(r_{ij} - r_0)^2 \quad (8)$$

To solve the equations of motion numerically, we use the Second Order Stochastic Runge-Kutta (SRKII) algorithm due to its high carefulness, compared by other presented methods. The Stochastic Runge-Kutta algorithm represents a solution of stochastic differential equations such as

$$\dot{x} = f(x) + g_w(t) \quad (7)$$

in which $g_w(t)$ is a Gaussian random number and relies in following two conditions:

$$\langle g_w(t) \rangle = 0,$$
$$\langle g_w(t)g_w(t') \rangle = 2D\delta(t-t'). \quad (8)$$

At the time $\Delta t$, $x$ is:

$$x_{\Delta t} = x_0 + \frac{(F_1 + F_2)\Delta t}{2} + (2D\Delta t)^{\frac{1}{2}}\psi \quad (9)$$

in which

$$F_1 = f(x_0),$$
$$F_2 = f(x_0 + F_1\Delta t + (2D\Delta t)^{\frac{1}{2}}\psi) \quad (10)$$

and $\psi$ is a Gaussian random number with zero mean, and unit variance [5].

We have put our carbon and hydrogen atoms in a 3-dimensional cubic initial network, and have let them move for a long time under the forces, and find their polymeric chain. Fig.1. shows a sample PE chains which contains many carbons each of which joined to two hydrogen atoms by a covalent bond.

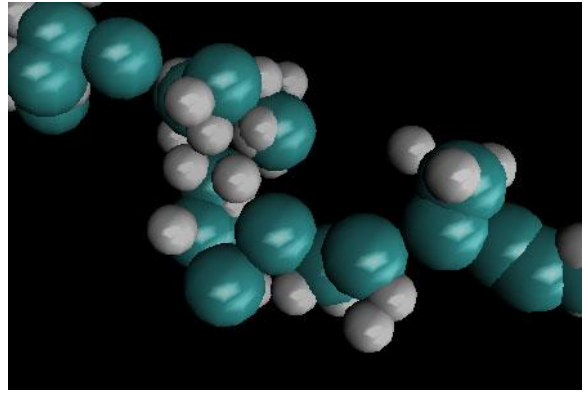

Fig. 1. *A PE sample with 25 Carbons in the backbone of the polymer each of which joining two Hydrogen atoms.*

**Our results and discussion**
To determine a polymer bending, we calculated Root Mean Square Displacement (RMSD) which is the mean end-to-end distance [16]

$$<r^2(t)>^{1/2} = <\sum_{d=1}^{3}[r_d(t) - r_d(0)]^2>^{1/2} \quad (11)$$

In which $r$ is position of the particle at step $t$, $d$ is the dimension, and the average is taken between many trials of an N-link polymer

$$<\xi> = \frac{\sum_{i=1}^{M}\xi_i}{M} \quad (2)$$

$M$ is number of all realizations.

RMSD denotes the overall size of our polymer in space. In PS, a hydrogen and 6 carbons of a phenyl group are attached to the backbone carbons of the polymer. Therefore, many sites are occupied by the surrounding groups. It means that there will be an important and significant excluded volume effect which has forbidden backbone carbons to insert into many sites.

In polyethylene, each carbon link to two hydrogen atoms, but in polystyrene, the phenyl group is linked to carbon atom alternatively, consequently movement and flexibility of polystyrene is lower than of them of polyethylene. In figure 2, plot of RMSD versus number of carbon atoms in backbone of these polymers is shown.

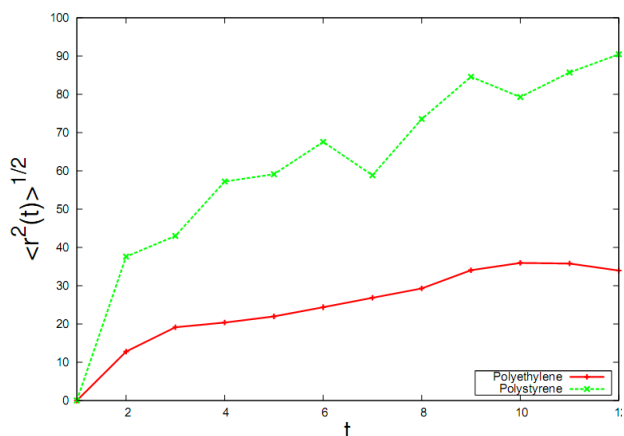

*Fig. 2. RMSD vs number of backbone Carbons at very low temperature of T=0.001 for PS and PE. Averaging is taken between 20 runs for each polymer each of which after 50000 time steps.*

Also in figure 3, the curve of RMSD versus temperature is depicted for polyethylene and polystyrene. As its shown in the plot, with increase in temperature, both density and stiffness increase. The high stiffness, $t_g$ and $t_m$ of polystyrene compared to polyethylene is because conformation energy of polystyrene is higher than polyethylene and rotation around c-c bond in polystyrene is harder than polyethylene.

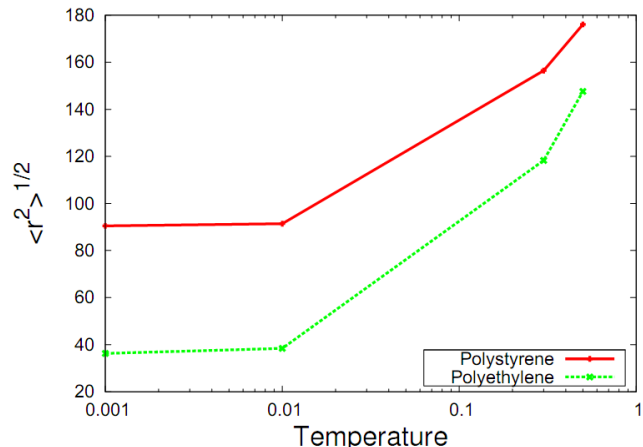

*Fig. 3. RMSD vs temperature for PS and PE. Each of the points are obtained by averaging between 20 polymers each of which running for 50000 time steps.*